\documentclass[traditabstract]{aa} % for the abstract without structuration 
\usepackage{graphicx}
\usepackage{txfonts}
\usepackage{natbib}
\usepackage{url}

\def\ltsima{$\buildrel<\over\sim$}
\def\lsim{\lower.5ex\hbox{\ltsima}~}
\def\gtsima{$\buildrel>\over\sim$}
\def\gsim{\lower.5ex\hbox{\gtsima}~}

\def\lir{$L_{\rm IR}$}

\def\micron{$\mu$m}
\def\hyperz{{\em Hyperz}}

\def\teff{\ifmmode T_{\rm eff} \else $T_{\mathrm{eff}}$\fi}

\def\lya{Ly$\alpha$} 

\def\ha{H$\alpha$}

\def\ebv{\ifmmode E_{B-V} \else $E_{B-V}$\fi}

\def\fesc{\ifmmode f_{\rm esc} \else $f_{\rm esc}$\fi}

\def\cm2{cm$^{-2}$}

\def\msunyr{M$_{\odot}$~yr$^{-1}$}
\def\lsun{L$_{\odot}$}
\def\msun{M$_{\odot}$}
\def\zphot{\ifmmode z_{\rm phot}\else$z_{\rm phot}$\fi}

\def\oiii{O{\sc iii}}

\def\nh{\ifmmode N_{\mathrm{HI}}\else $N_{\mathrm{HI}}$\fi}
\def\nhi{\ifmmode N_{\mathrm{HI}}\else $N_{\mathrm{HI}}$\fi}

\def\vexp{\ifmmode v_{\rm exp} \else v$_{\rm exp}$\fi}
\def\taua{\ifmmode \tau_{a}\else $\tau_{a}$\fi}

\begin{document}
   \title{Far infrared spectral energy distributions of two very high redshift dropout candidates}
\title{Far infrared constraints on the contamination by dust obscured galaxies of high-z dropout searches}
\titlerunning{contamination by dust obscured galaxies of high-$z$ dropout searches}
   %\subtitle{I. Overviewing the $\kappa$-mechanism}

   \author{F. Boone
          \inst{1,2}
          \and
          D. Schaerer
          \inst{3,2}
          \and
          R. Pell\'o
          \inst{1,2}
          \and
          D. Lutz
          \inst{4}
          \and
          A. Weiss
          \inst{5}
          \and
          E. Egami
          \inst{6}
          \and
          I. Smail
          \inst{7}
          \and
          M. Rex
          \inst{6}
          \and
          T. Rawle
          \inst{6}
          \and
          R. Ivison
          \inst{8,9}
          \and
          N. Laporte
          \inst{1,2}
          \and
          A. Beelen
          \inst{10}
          \and
          F. Combes
          \inst{11}
          \and
          A. W. Blain
          \inst{12}
          \and
          J. Richard
          \inst{13}
          \and
          J.-P. Kneib
          \inst{14}
          \and
          M. Zamojski
          \inst{3}
          \and
          M. Dessauges-Zavadsky 
          \inst{3}
          \and
          B. Altieri
          \inst{15}
          \and
          P. van der Werf
          \inst{16}
          \and
          M. Swinbank
          \inst{7}
          \and
          P. G. P{\'e}rez-Gonz{\'a}lez
          \inst{17}
          \and
          B. Clement
          \inst{14}
          \and
          R. Nordon
          \inst{4}
          \and
          B. Magnelli
          \inst{4}
          \and
          K. M. Menten
          \inst{5}
          }

   \institute{Universit\'e de Toulouse; UPS-OMP; IRAP;  Toulouse, France
            \and
            CNRS; IRAP; 9 Av. colonel Roche, BP 44346, F-31028 Toulouse cedex 4, France
            \and
             Geneva Observatory, Universit\'e de Gen\`eve, 51 chemin des Maillettes, 1290 Versoix, Switzerland 
             \and
             Max-Planck-Institut f\"ur extraterrestrische Physik,  Postfach 1312, 85741 Garching, Germany 
             \and
            Max-Planck-Institut f\"ur Radioastronomie, Auf dem H\"ugel 69, 53121 Bonn, Germany
            \and
            Steward Observatory, University of Arizona, 933 North Cherry Avenue, Tucson, AZ 85721, USA
            \and
            Institute for Computational Cosmology, Durham University, South Road, Durham DH1 3LE, UK
            \and
            UK Astronomy Technology Centre, Science and Technology Facilities Council, Royal Observatory, Blackford Hill, Edinburgh EH9 3HJ, UK 
            \and
            Institute for Astronomy, University of Edinburgh, Blackford Hill, Edinburgh EH9 3HJ, UK
            \and
            Institut d'Astrophysique Spatiale, bat 121, Universit\'e Paris Sud 11 \& CNRS (UMR8617), 91405 Orsay Cedex, France
            \and
            LERMA, Observatoire de Paris, 61 avenue de l'Observatoire, 75014 Paris, France
            \and
            California Institute of Technology, 1200 East California Boulevard, Pasadena, California 91125, USA
            \and
            Centre de Recherche Astrophysique de Lyon, Universit\'e Lyon 1, 9 Avenue Charles Andr\'e, F-69561 Saint Genis Laval, France
            \and
            Laboratoire d’Astrophysique de Marseille, CNRS- Universit\'e Aix-Marseille, 38 rue F. Joliot-Curie, 13388 Marseille Cedex 13, France
            \and 
            Herschel Science Centre, European Space Astronomy Centre, ESA, Villanueva de la Cañada, 28691 Madrid, Spain
            \and
            Leiden Observatory, Leiden University, P.O. Box 9513 , NL-2300 RA Leiden, The Netherlands 
            \and
Departamento de Astrof\'{\i}sica, Facultad de CC. F\'{\i}sicas, Universidad Complutense de Madrid, 28040 Madrid, Spain
             }

   \date{Received date; accepted date}

% \abstract{}{}{}{}{} 
% 5 {} token are mandatory
 
  \abstract
   {The spectral energy distributions (SED) of dusty galaxies at intermediate redshift may look similar to very high redshift galaxies in the optical/near infrared (NIR) domain. This can lead to the contamination of high redshift galaxy searches based on broad band optical/NIR photometry by lower redshift dusty galaxies as both kind of galaxies cannot be distinguished. The contamination rate could be as high as 50\%. { This work shows how the far infrared (FIR) domain can help to recognize likely low-z interlopers in an optical/NIR search for high-z galaxies.}
We analyse the FIR SEDs of two galaxies proposed as very high redshift ($z>7$) dropout candidates based on deep Hawk-I/VLT observations. The FIR SEDs are sampled with PACS/{\em Herschel} at 100 and 160\,$\mu$m, with SPIRE/{\em Herschel} at 250, 350 and 500\,$\mu$m and with LABOCA/APEX at 870\,$\mu$m. { We find that redshifts $>7$ would imply  extreme FIR SEDs (with dust temperatures $>100$\,K and FIR luminosities $>10^{13}$\,$L_{\odot}$). At z$\sim$2, instead, the SEDs of both sources would be compatible with that of typical ULIRGs/SMGs}. Considering all the data available for these sources from visible to FIR we re-estimate the redshifts and we find  $z\sim$1.6--2.5. Due to the strong spectral breaks observed in these galaxies,
standard templates from the literature fail to reproduce the
visible-near IR part of the SEDs even when additional extinction is
included. These sources resemble strongly dust obscured galaxies selected in {\em Spitzer} observations with extreme visible-to-FIR colors, and the galaxy GN10 at $z=4$. 
Galaxies with similar SEDs could contaminate other high redshift surveys.
}

   \maketitle

%%%%%%%%%%%%%%%%%%%%%%%%%%%%%%%%%%%%%%%%%%%%%%%%%%%%%%%%%%%%%%%%%%%%%%%%%%%%%%%
\section{Introduction}

 Observing galaxies up to very high redshifts allows us to study directly the formation and evolution of structures in the expanding Universe. Finding galaxies at ever higher redshifts has therefore become one of the main areas of extragalactic astronomy. The most common technique is to use known broad features in the spectral energy distributions (SEDs) of galaxies to identify high redshift sources in deep  optical and near infrared (NIR) multi-band observations. In particular the Lyman break is widely used to select sources by redshift noting their disappearance in bands below a given wavelength, the so-called drop-out technique \citep{1996ApJ...462L..17S}. With this technique and state of the art telescopes and instruments it is now possible to select sources that are good candidates for being at the end or within the epoch of reionization \citep{2006A&A...456..861R, 2009ApJ...697.1907Z, 2010MNRAS.403..960M, 2010MNRAS.403..938W, 2010ApJ...709L..16O, 2010ApJ...709L.133B, 2010ApJ...725.1587B}. 

Low redshift galaxies, however, can have very steep SEDs resembling a break in the UV/optical/NIR.  This can lead to contamination of the dropout selection of very high-z galaxies, and hence to
erroneous estimates of the star formation rate density, stellar masses, and others, although these
effects are currently difficult to quantify.
Such objects have been found and discussed by several authors 
\citep[see e.g.][]{2000ApJ...531..624D,2005ApJ...635..832M,Schaerer07ero,2007MNRAS.376.1054D,
2007ApJ...665..257C,Capak11}.
Confirming the photometric redshifts of high-z galaxies by identifying spectral lines is challenging because the sources are generally too faint for spectroscopic follow up observations or because they may intrinsically lack
\lya\ emission \citep[but see][]{2010arXiv1011.5500V}.
%\citep[but see][]{2010arXiv1011.5500V,2010Natur.467..940L}.

The recent developments of space far infrared instrumentation offer new perspectives in this domain. In particular, with the advent of the {\em Herschel} Space Observatory it is now possible to sample the Far Infrared (FIR) SEDs of the galaxies, where the thermal dust emission dominates. The shape of the FIR SED universally looks like a broad bump which can be used to further constrain the optical/NIR photometric redshifts. Although the wavelength of the FIR SED peak also depends on the dust temperature, the limited range of average temperatures  observed so far in galaxies { \citep[between 20 and 60\,K averaged over the entire galaxies, see, e.g.,][]{2006ApJ...650..592K,2010A&A...518L..28M,2010MNRAS.409...22M, Wardlow10}} can be used as a prior and makes it possible to discriminate between intermediate ($z<3$) and very high redshifts ($z>6$).

FIR observations of high-z candidates are also essential to characterize their star forming and dust properties and thus interpret correctly their contribution to the cosmic history of star formation and reionization.

Recently \citet{Laporte11} identified ten $z>7$ candidates in the field of the cluster Abell 2667 using photometric dropout criteria based on deep observations with HAWK-I on the ESO Very Large Telescope (VLT). { Comparing to other studies and in particular to the WIRCAM Ultra Deep Field Survey (WUDS; Pello et al in prep), which is based on deeper optical observations bluewards to the I-band}, they estimated that 50-75\% of these candidates could in fact be lower redshift interlopers. Here, we study two galaxies of this sample that are clearly detected by {\em Herschel}, namely the sources named 'z1' and 'Y5'. Our goal is to determine whether they could be interlopers and to understand their nature.
%among the ten sources of the \citet{Laporte11} sample. 
The redshift probability distributions of these two sources  derived by \citet{Laporte11} from SED fitting to deep Optical/NIR photometry show a prominent peak at $z=$7.6 and 8.6 respectively. However, a secondary peak at lower redshift around $z\sim 2$, indicates that they also could be interlopers. \citet{Laporte11} also noted that the 24\,$\mu$m detection of z1 with MIPS/{\em Spitzer} (Y5 is out of the Spitzer map) seems difficult to reconcile with the high-z solution.
% withThis is particularly the case for Y5 when prior luminiosity functions are taken into account in the fitting process  \citep[see ][ for details]{Laporte11}.
We use new {\em Herschel} and LABOCA observations of Abell\,2667, to reconstruct the FIR part of their SEDs. We can thus further constrain their redshifts and study their physical properties.

The layout of the article is as follows: Section\,\ref{sec:obs} gives a presentation of the observations and data analysis. In Section\,\ref{sec:fir}, the FIR part of the SEDs is analyzed. In Section\,\ref{sec:multiw}, the complete SEDs are used to estimate the redshifts and discuss the physical properties of the two galaxies. In Section\,\ref{sec:comparison}, we compare the two galaxies to other similar galaxies found in the literature. Section\,\ref{sec:conclusion} gives the conclusions. 
We assume a $\Lambda$-cosmology with $H_0=70$\,km\,s$^{-1}$\,Mpc$^{-1}$, $\Omega_M=0.3$ and $\Omega_{\Lambda}=0.7$.

%%%%%%%%%%%%%%%%%%%%%%%%%%%%%%%%%%%%%%%%%%%%%%%%%%%%%%%%%%%%%%%%%%%%%%%%%%%%%%%
\section{Observations and Data Analysis}\label{sec:obs}

\subsection{Observations and reduction}
{\em Herschel} observations were obtained in the framework of the {\em Herschel} Lensing Survery (HLS) described by \citet{2010A&A...518L..12E}. They include PACS data at 100 and 160\,$\mu$m, and SPIRE observations at 250, 350 and 500\,$\mu$m. The data reduction was done with the HIPE software as described by \citet[][]{2010A&A...518L..13R} and \citet{2010A&A...518L..14R}.  

The large APEX Bolometer Camera \citep[LABOCA][]{2009A&A...497..945S} is a bolometer array  operating at 870\,$\mu$m and mounted on the APEX telescope in the desert of Atacama, Chile \citep{2006A&A...454L..13G}. The LABOCA observations were conducted during the summer 2010. The cluster was mapped in spiral mode during 30 hours covering a circular field of $\sim$6$'$ in radius. The data were reduced with the BoA\footnote{\url{http://www.apex-telescope.org/bolometer/laboca/boa/}} software. The noise is not uniform over the map and the RMS is in the range 1.1--3.0\,mJy, the highest values being reached at the edges of the map.

We also obtained a VLA  1.4\,GHz continuum map of Abell 2667 (PI: R. Ivison) with an RMS of 46\,$\mu$Jy.

\subsection{Analysis}
The astrometry of all the maps was corrected to  align them with the VLT Ks image.
All the optical drop-out sources of \citet{Laporte11} were inspected in the {\em Herschel} and LABOCA images. 
Two of them, z1 and Y5, are detected in several FIR bands. As IRAC/{\em Spitzer} and MIPS/{\em Spitzer} data are available for z1, the source can be followed from one band to the  next one by increasing wavelength despite the decreasing resolution. Its identification is therefore robust. 

{ For Y5 there is a larger gap in the SED due to the lack of data between 8\,$\mu$m and 100\,$\mu$m, and due to the fact that it lies at the noisy edges of the 100 and 160\,$\mu$m maps, where it is not detected. 
However, Y5  is the only source detected at  4.5\,$\mu$m within a radius of 3$''$  (i.e., $\sim1/3^{\rm rd}$ of the 250\,$\mu$m beam radius) around the 250\,$\mu$m peak, its identification with the SPIRE detection is therefore very likely.}

The fluxes are measured at the positions of the two galaxies by PSF fitting in apertures with a radius equal to FWHM$/3$, where FWHM is the PSF full width at half maximum, i.e., 5.6$''$, 11.3$''$, 18.1$''$, 24.9$''$, 36.6$''$ and 22.5$''$ from 100 to 870\,$\mu$m. The last (LABOCA) FWHM corresponds to the APEX beam convolved by a gaussian of 12$''$. The sources were deblended from the neighbouring sources by subtracting PSFs at the positions of the neighbours derived from the 250\,$\mu$m map. Observations at these wavelengths with these resolutions are affected by source confusion. As a consequence a measured flux cannot be directly interpreted as the true flux of a single underlying source. A correct treatment of the effect of source confusion on flux measurements (a.k.a. flux 'deboosting') requires a prior knowledge of the source counts towards low fluxes at the given wavelength. We followed the method presented by \citet{2010ApJ...718..513C} based on a Bayesian analysis. For the prior source counts we extrapolated toward low fluxes the results of  \citet{2010A&A...518L..30B} for PACS bands, \citet{2010A&A...518L..21O} for SPIRE bands and \citet{2006MNRAS.372.1621C} for the LABOCA band. 
%In principle the problem should be solved for all bands simultaneously but such an analysis is beyond the scope of this paper and we solve each band independently. 

 Blending affects z1 photometry at $\lambda\ge 250$\,$\mu$m and Y5 photometry at $\lambda\ge 500$\,$\mu$m.
%Only z1 photometry is affected by blending at $\lambda>250$\,$\mu$m. 
And the effect of deboosting is small ($<20\%$), except for the 870\,$\mu$m  measurement of Y5 that corresponds to a 2.6$\sigma$ signal and that we chose to consider as a tentative detection. The deboosted flux of Y5 at  870\,$\mu$m is 1.8$\pm$1\,mJy for a measured flux of 2.5$\pm$0.95\,mJy. 

\begin{figure*}
\resizebox{\textwidth}{!}{
 \includegraphics[]{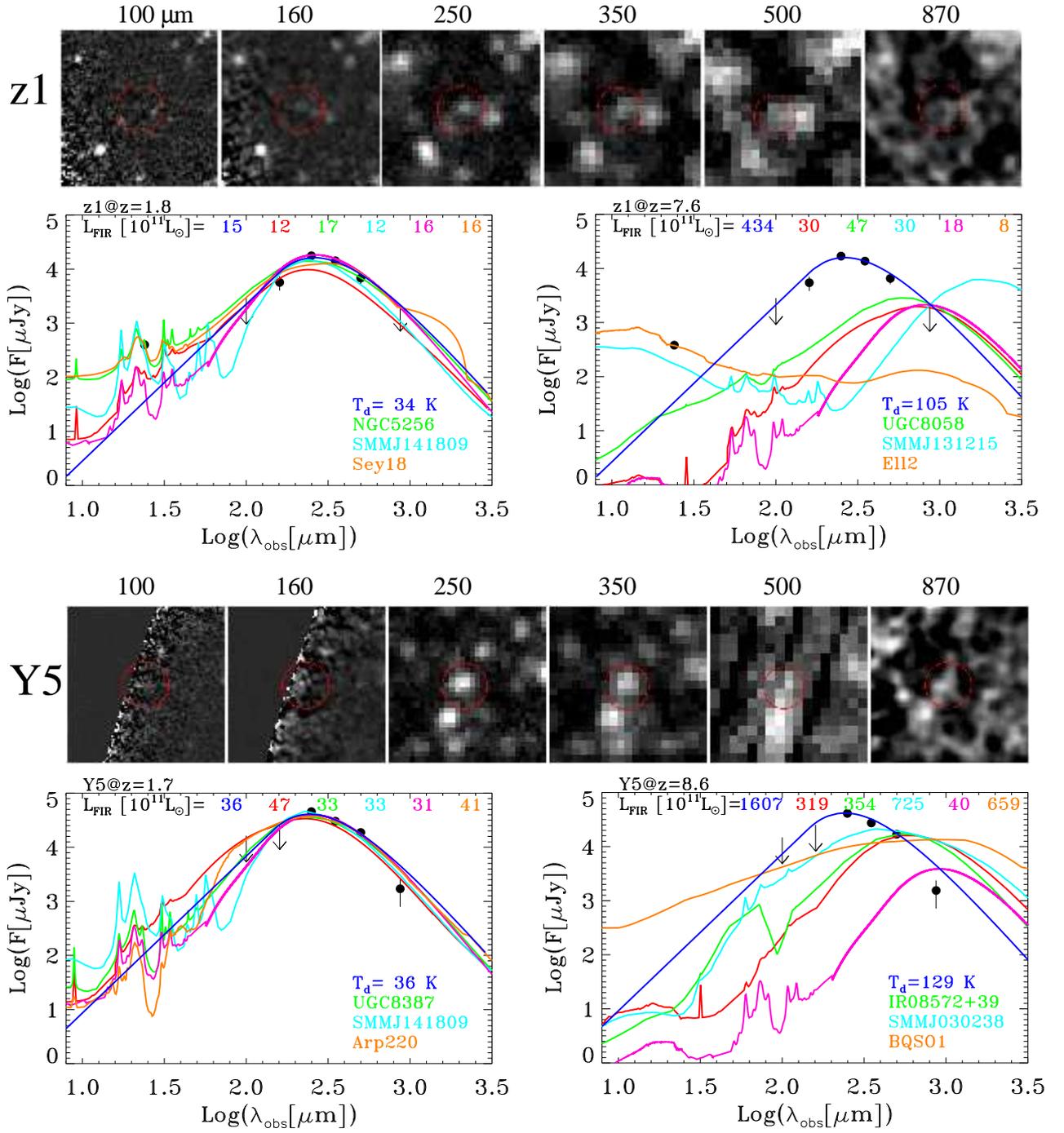}
}
\caption{The top figures show the postage stamps in the five {\em Herschel} bands (100, 160, 250, 350, 500 $\mu$m from left to right) and the LABOCA band (870 $\mu$m, the right-most stamp) centered on z1 and the best fit SEDs for the low redshift (left) and high redshift (right) solutions. The bottom figures show the same for the source Y5. The red circle in the postage stamps is meant to guide the eye, it has a radius of 27$''$ and it is centered on the source position. 
The blue curves correspond to the best fit modified black body SEDs with the parameters written in blue, the magenta curves correspond to the SMM\,J2135-0102 model \citep{2010Natur.464..733S,2010A&A...518L..35I}, the other curves correspond to the best fit templates of the \citet{2001ApJ...556..562C} library (red), the \citet{2008A&A...484..631V} library (green), the \citet{2010A&A...514A..67M, 2010ApJ...712..942M} library (cyan) and the \citet{Polletta07} library (orange). The names of the best fit templates as well as $L_{\rm FIR}$ are written on the figure with the same color codes.
}
\label{fig:stampsseds}
\end{figure*}

None of the two sources are detected in the VLA map.
The measured FIR fluxes of the sources as well as their optical to near-IR photometry from \citet{Laporte11} are listed in Table\,\ref{tab:fluxes}. 
Postage stamps of the {\em Herschel} and LABOCA bands centered at the source positions as well as FIR SED fits are shown in the Fig.\,\ref{fig:stampsseds}. Y5 is close to the border of the PACS maps where the noise is higher, hence the high upper limits.

\begin{table}

\caption{Multi-wavelength SED of z1 (col 3) and Y5 (col 4). Optical and IR photometry (rows 1-10) is taken from 
\citet{Laporte11}. Upper limits are 3 $\sigma$.
Rows 12-17 give the {\em Herschel} and LABOCA source flux. When the measured flux 
(not the estimated flux) is $<2.5 \, \sigma$ the $3 \sigma$ value is given as an upper limit.
No entry indicates the lack of data. All fluxes are given in milli-Jansky.}\label{tab:fluxes}
\begin{tabular}{lrrr}
Band/instrument & $\lambda_{\rm eff}$ [\micron] & z1 & Y5 \\
%Band & effective wavelength [\micron] & z1 & Y5 \\
\hline
I           & 0.79 & $<$3.6e-5           & $<$3.6e-5 \\
z           & 0.92 & $<$1.4e-4           & $<$1.4e-4 \\
Y           & 1.02 & (2.0$\pm$0.3)e-4    & $<$6.3e-5 \\
J           & 1.26 & (1.75$\pm$0.05)e-3  & (6.9$\pm$0.5)e-4   \\
H           & 1.63 & (1.33$\pm$0.07)e-3  & (9.3$\pm$0.03)e-4  \\
Ks          & 2.15 & (2.29$\pm$0.06)e-3  & (1.77$\pm$0.07)e-3 \\ 
IRAC        & 3.6  & (6.98$\pm$0.06)e-3  & (2.91$\pm$0.11)e-3 \\
            & 4.5  & (10.20$\pm$0.09)e-3 & (3.73$\pm$0.10)e-3 \\
            & 8.0  & (9.91$\pm$0.96)e-3  &  \\
MIPS        & 24   & 0.340$\pm$0.040 &  \\   
PACS & 100 &  $<$3.3        & $<$18 \\
& 160 &  6.3$\pm$2.0   & $<$30 \\ 
SPIRE & 250 & 19.4$\pm$1.6   & 45.5$\pm$1.6 \\
& 350 & 15.7$\pm$1.4   & 30.3$\pm$1.2 \\
& 500 &  7.5$\pm$1.7   & 19.4$\pm$2.1 \\
LABOCA & 870 &  $<$2.5        & 1.8$\pm$1.0  \\
VLA & 2.1$\times10^5$ & $<$0.14  & $<$0.14 \\
\end{tabular}
\end{table}

%%%%%%%%%%%%%%%%%%%%%%%%%%%%%%%%%%%%%%%%%%%%%%%%%%%%%%%%%%%%%%%%%%%%%%%%%%%%%%%
\section{Analysis of the FIR SEDs}\label{sec:fir}

The following models or templates have been fitted to the FIR measurements (cf.\ Fig.\ \ref{fig:stampsseds}):
\begin{itemize} 
\item A modified black body SED parameterized as described by \citet{2003MNRAS.338..733B}, with emissivity fixed to $\beta=1.5$ and the Wien correction parameter $\alpha=2.9$. { These values are adapted to submillimeter galaxies \citep[SMGs; ][]{2005ApJ...622..772C} and local ultraluminous infrared galaxies \citep[ULIRGs; ][]{2000MNRAS.315..115D, 2003MNRAS.338..733B}}. The free parameters are the total FIR luminosity, $L_{\rm FIR}$,  defined as the luminosity emitted in the range 8--1000\,$\mu$m, and the dust temperature, $T_{\rm d}$.
\item { The 105 galaxy templates built by \citet{2001ApJ...556..562C}. The templates are fitted without rescaling.} \footnote{We found that such an additional scaling parameter was not required to obtain good fits.}
\item ULIRG templates built by \citet{2008A&A...484..631V}, with a scaling parameter, $L_{\rm FIR}$.
\item { The starburst, Seyfert, and active galactic nuclei (AGN) templates of \citet{Polletta07}, with a scaling parameter, $L_{\rm FIR}$.}
\item Templates built by \citet{2010A&A...514A..67M, 2010ApJ...712..942M} to fit high redshift galaxies with detected but poorly sampled submm emission, with a scaling parameter, $L_{\rm FIR}$. 
\item The SED fit to the observations of SMM\,J2135-0102 \citep{2010Natur.464..733S,2010A&A...518L..35I} with a scaling parameter, $L_{\rm FIR}$.
\end{itemize}

The fit is performed by finding the maximum likelihood assuming gaussian probability distributions for the measurements. When there is no detection the 3-$\sigma$ value is used as a hard upper limit, i.e., the probability is assumed to be uniform in the [0, 3$\sigma$] interval and zero outside. The redshifts are fixed to the solutions derived by \citet{Laporte11} from the optical/NIR photometry, i.e., $z=$1.8 and 7.6 for z1 and $z=$1.7 and 8.6 for Y5. The corresponding magnification factors are  $\mu=$1.12 and 1.17  for  z1 and $\mu=$1.04 and 1.15 for Y5.
The MIPS/{\em Spitzer}  24\,$\mu$m flux of z1 (Y5 has no 24\,$\mu$m data available) was taken into account to fit various galaxy templates, but ignored to 
fit the modified black body since it is most likely dominated by polycyclic aromatic hydrocarbons (PAHs).

For both sources we find reasonable fits at low redshift for the modified black body and the various galaxy templates. { A ULIRG template from the \citet{2008A&A...484..631V} library, a submillimeter detected galaxy template from the \citet{2010A&A...514A..67M, 2010ApJ...712..942M} library and a Seyfert template from the \citep{Polletta07} library are able to reproduce the 24\,$\mu$m emission of z1}. The modified black body model gives dust temperatures of 34 and 40\,K, for z1 and Y5 respectively, which are typical values for integrated dust temperatures in LIRGs. The infrared luminosities, $L_{\rm FIR}$, are in the range $(1.2-1.7)\times10^{12}$\,L$_{\odot}$  and $(3.1-4.7)\times10^{12}$\,L$_{\odot}$ for z1 and Y5 respectively. There is a noticeable agreement in $L_{\rm FIR}$ between the modified black body model and the various templates. These galaxies would therefore be typical ULIRGs/SMGs at $z\sim 2$. This is consistent with the general picture of galaxy evolution now widely observed, i.e., that the contribution of ULIRGs to the cosmic SFR is expected to peak  at $z\sim 2$ where it should be comparable to that of the more 'normal' galaxies \citep[see e.g. ][]{2011arXiv1102.3920M}.

For the high redshift solutions ($z>7.5$), instead, the modified black body requires for both sources very high dust temperatures, i.e., 105 and 129\,K for z1 and Y5, respectively. Such high temperatures averaged over an entire galaxy are extreme. \footnote{ To our knowledge there was only one such extreme case reported so far: the host of the lensed quasar APM08279 at $z=3.9$ \citep{2007A&A...467..955W,2009ApJ...690..463R} . It requires two dust components at 75 and 220\,K and the highest temperatures most likely result from heating by the powerful quasar}. This can be seen from the impossibility to find any good fit in the different template libraries, which were built from observed galaxies. However, the dust properties of galaxies at such high redshifts are unknown and dust temperatures above 100\,K cannot be ruled out. This would imply that the FIR luminosities are of the order 0.5 and 1.5\,$\times10^{14}$\,L$_{\odot}$ for z1 and Y5, respectively, i.e. both sources would be classified as Hyper Luminous Infrared Galaxies (HyLIRG). While the nature of HyLIRGs is still a matter of debate \citep[e.g.\ ][]{2010A&A...515A..99R} and their density at very high redshift is not well known, they are extreme sources with a lower number density than ULIRGs.

The radio continuum upper limits are too high to constrain the SED fitting. We find that for Y5 only at the low redshift solution ($z\sim 2$) the Chary \& Elbaz template is close to the 3$\sigma$ upper limit.

Thus, in summary, by comparing the FIR photometry to known galaxy SEDs and by taking into account the expected temperature and luminosity range of high redshift galaxies, the very high redshift solutions derived from the optical/NIR photometry seem to be less likely than the low redshift solution. The two sources are most likely typical ULIRGs at $z\sim 2$. 
This result puts strong constraints on the optical/NIR analysis which gave a much higher probability to the very high redshift solution when no prior luminosity function was taken into account. 
The FIR data alone, however, cannot be used to derive any accurate photometric redshift because of the redshift-temperature degeneracy.

%%%%%%%%%%%%%%%%%%%%%%%%%%%%%%%%%%%%%%%%%%%%%%%%%%%%%%%%%%%%%%%%%%%%%%%%%%%%%%%
\section{Analysis of the complete SED from visible to FIR}\label{sec:multiw}
\label{s_models}
We will now examine all the data from the visible to the FIR ranges to improve the redshift estimate of our galaxies
and to examine the nature and physical properties of these sources.

\subsection{Method}
To model the SED of the two sources we use a modified version of the \hyperz\
photometric redshift code of \citet{hyperz} described in \citet{SdB09}. 
{ Non-detections are treated as the usual case “1” of Hyperz, i.e.\ 
the flux in these filters is set to zero, with an error bar corresponding to the flux at 1$\sigma$ level.}
The basic spectral templates are taken from the Bruzual \& Charlot models
\citep{BC03}, computed for a variety of star-formation histories and
metallicities.
{ Although applicable only to a limited part of the spectrum, we use
these templates here to constrain redshift, extinction, and stellar mass
in particular.}
For the Bruzual \& Charlot templates we consider variable extinction
with $A_V$ up to 8 magnitudes for the \citet{Calzetti00} attenuation law.
We have also explored other extinction laws.

The code, initially designed to fit rest frame UV to near-IR (stellar) emission, 
can also easily be used to include the thermal mid-IR and beyond.
To cover the entire spectral range from the visible to the millimeter domain,
and to compare our sources with SEDs of very different galaxy types,
we have compiled a large variety of spectral templates from
the GRASIL models of \citet{Silva98}, the library of \cite{2001ApJ...556..562C},
\citet{2009ApJ...692..556R},
the starburst, Seyfert, and AGN templates of \citet{Polletta07},
%revds
the ULIRG templates of  \citet{2008A&A...484..631V},
the sub-mm galaxy templates of \citet{2010A&A...514A..67M},
and the model fit to SMM\,J2135-0102 \citep{2010Natur.464..733S,2010A&A...518L..35I}.
 Extinction can also be added to these spectral templates; SED fits with and
without additional extinction will be discussed below.

 We have carried out both fits of the entire SED (optical, near-IR, and IR) and 
fits up to 8 \micron\ only { (for the Bruzual \& Charlot templates)}.
For each template set the free parameters are redshift and (additional) $A_V$.
Physical parameters such as the infrared luminosity, \lir, defined as the luminosity emitted in the range 2-1000 \micron; the IR star-formation rate, SFR; and the stellar mass, are subsequently derived
from the best-fit templates. In contrast to the IR fits discussed in 
Sect.\ \ref{sec:fir} we have no handle on the dust temperature, since this is 
not a parameter describing the SEDs used here.
We have also checked that the two independent fitting methods used here 
and in Sect.\ \ref{sec:fir} give consistent results.

\begin{figure*}[tb]
{\centering
\includegraphics[width=8.8cm]{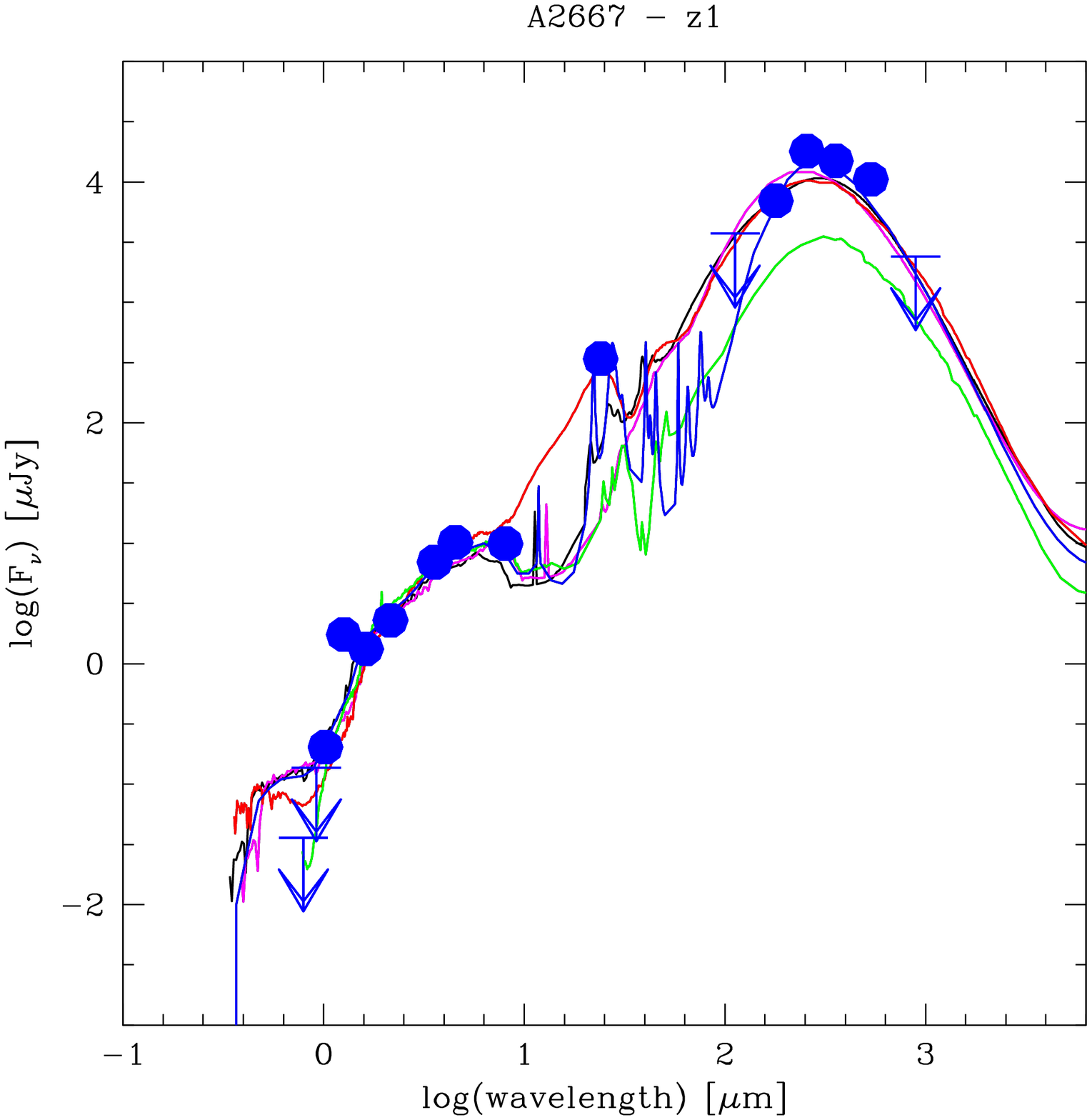}
\includegraphics[width=8.8cm]{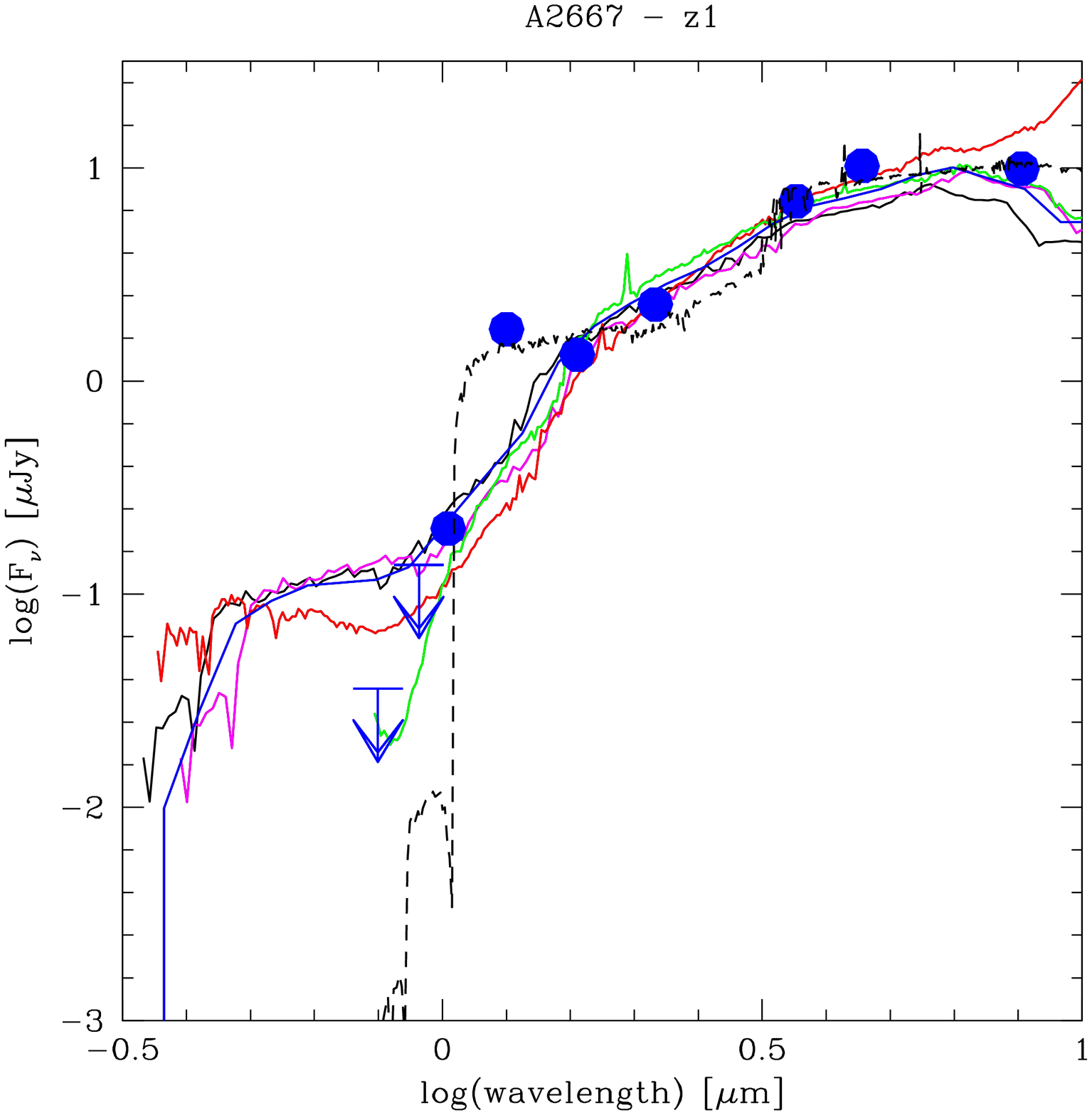}
}

\caption{Fits to the observed SED of source z1 (photometry shown by blue symbols, including 3 $\sigma$ upper limits) 
using different spectral templates: \citet{2001ApJ...556..562C} (black = best-fit template, and magenta = template with maximum
IR luminosity), \citet{Polletta07} (red  = global best-fit template, and green = best-fit to visible--near-IR SED excluding
the thermal IR), and \citet{2010A&A...514A..67M} (blue). 
The best-fit SED with the templates of Vega et al.\ (2008), very similar to the one using Polletta's templates, is not 
shown here for simplicity. 
{\bf Left:} Global visible to sub-mm SED. {\bf Right:} Zoom on visible to near-IR part of the SED including for
comparison also the best-fit SED at high redshift ($z=7.5$) from \citet[][dashed line]{Laporte11}, which is 
most likely excluded due to our Herschel detections.}
\label{fig_z1}
\end{figure*}

\begin{figure*}[tb]
{\centering
\includegraphics[width=8.8cm]{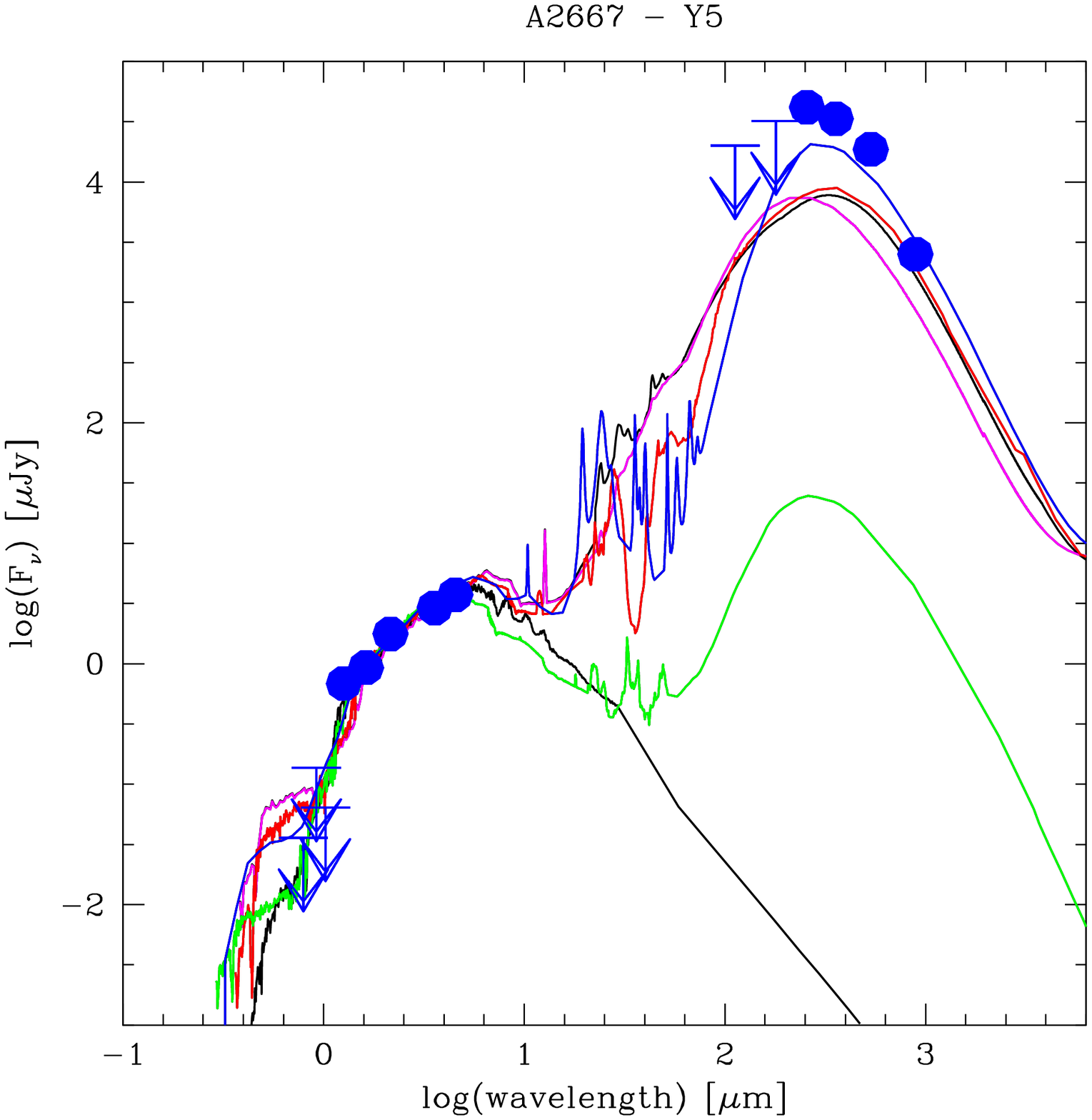}
\includegraphics[width=8.8cm]{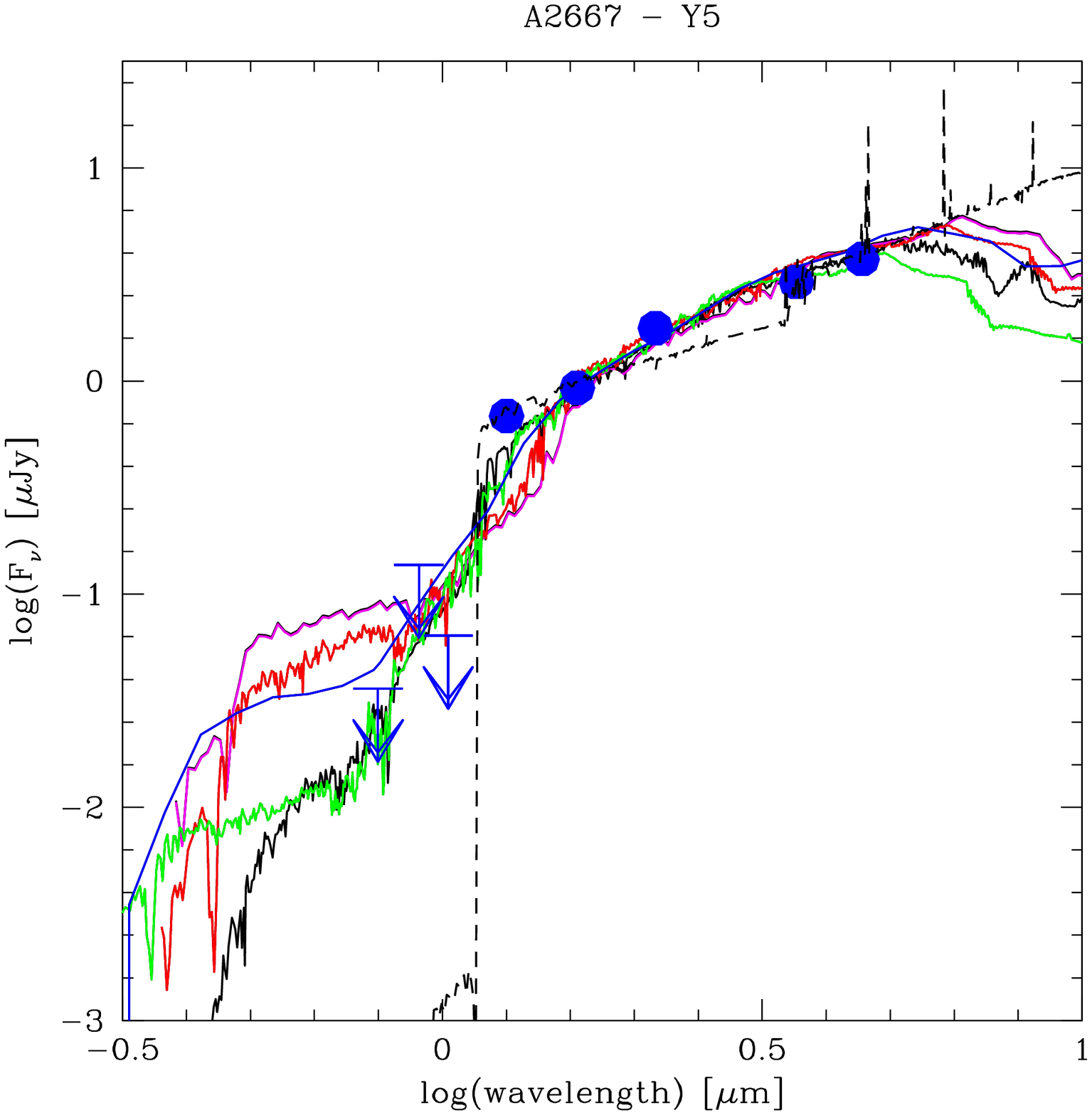}
}
\caption{Same as Fig.\ \protect\ref{fig_z1} for Y5. The best-fit SED at high redshift (black dashed line) is for $z=8.28$.}
\label{fig_y5}
\end{figure*}

\subsection{Photometric redshifts}
 As discussed in depth by \citet{Laporte11} the best-fit photometric redshifts
of our sources derived from the optical to near-IR photometry (up to 8 \micron)
and using standard spectral templates is consistently found at $z>7$ with a lower probability at low $z$. 
This result remains unchanged with the exploration of a wider range of
extinction, different attenuation/extinction laws, and templates sets used here
compared to \citet{Laporte11}
On the other hand, analysis of the IR SED and other arguments
clearly favour low redshifts ($z \sim$ 1.5--2.5) as discussed above.
Subsequently we therefore limit ourselves to $z<4$ and attempt to refine
the photometric redshift of the two sources.

\subsection{Results for z1}
Overall the global, multi-wavelength SED fits for this source are rather satisfactory, as shown in Fig.\ \ref{fig_z1},
albeit there are important discrepancies in the optical domain (cf.\ below).
%revds
Several templates (i.e.\ Polletta's, the ULIRGs of \citet{2008A&A...484..631V}, and the best-fit SMG template from \citet{2010A&A...514A..67M})
also reproduce the 24 \micron\ flux, and the observed 100 and 870 \micron\ fluxes are within 2--3 $\sigma$
of the model.
Interestingly, the best fits for both sets of Polletta's and Michalowski's templates are
found with templates for active galaxies.

The best-fit redshift found with these templates are between $z \approx 2.24$ and 2.57.
The resulting IR luminosity  is
$L_{\rm IR} \sim (2.6-3.2) \times 10^{12}$ \lsun, the corresponding SFR $\approx$
450--550 \msunyr\ using the standard \citet{Kennicutt98} calibration.
Fits to the IR part with the SMM\,J2135-0102 template yield $\zphot \approx 2.0$.
A somewhat lower redshift of $z \approx 1.7$ is found with \citet{BC03} templates using the SED 
up to the IRAC bands. The estimated extinction is $A_V \sim 2.6$, the stellar mass 
$M_\star \sim 6\times 10^{10}$ \msun\ for the same Salpeter IMF adopted by \citet{Kennicutt98}. 
However, these values should be taken with caution as the fits are not of good quality.
For comparison from the absolute H-band magnitude ($M_B \approx -23.0$) one obtains
$M_\star \sim 3\times 10^{10}$ \msun\ using the mass-to-light ratio adopted by \citet{Wardlow10}
for SMGs.

At a more detailed level (see right panel), all spectral templates have some difficulty to reproduce
the steep, observed SED between the visible (I, z bands) and the near-IR (Y and J here), and they
predict a flux excess in the optical domain. Below we will show that this also holds when
variable extinction is added to the empirical templates.
The same is also true for all other templates we have examined, including the theoretical galaxy templates 
of \citet{BC03}. This sharp drop is of course the reason why this source was selected as an optical 
dropout (Y-drop).

\subsection{Results for Y5}
For this source the global fits 
are less good than for z1. This is due 
to the fact that Y5 shows a higher flux ratio between the thermal-IR and the near-IR than z1, 
whose SED already required  templates with extreme IR/near-IR fluxes.
For example, the Arp 220 template from \citet{Polletta07}, shown in red, underpredicts the
IR flux by a factor $\ga 5$. The only template coming near the observed IR emission is from 
the SMG library of \citet{2010A&A...514A..67M} (SMMJ221725.97+001238). 
With a best-fit redshift of $\zphot \approx 2.15$ this translates to $L_{\rm IR} = 2.2 \times 10^{12}$ \lsun, 
corresponding to SFR $\approx$ 380 \msunyr.
Fits to the IR part with the SMM\,J2135-0102 template yield $\zphot \approx 1.8$.
A best-fit redshift of $z \approx 1.95$ is found with \citet{BC03} templates using the SED 
up to the IRAC bands. The estimated extinction is $A_V \sim 1.6$, the stellar mass 
$M_\star \sim 3\times 10^{10}$ \msun. 
 However, these values should be taken with caution as the fits 
in the domain close to the optical are not of good quality.
Again, using the absolute H-band magnitude ($M_B \approx -22.3$) one obtains
$M_\star \sim 2\times 10^{10}$ \msun\ with the assumptions already mentioned above.

As for z1, the visible--near-IR drop of the SED (see right panel) is not well fit by the 
spectral templates, predicting that the source should be detectable in the
visible (I, z, Y bands in particular), in contrast to our observations.
The template fitting best this part of the spectrum is an S0 template from
\citet{Polletta07}, shown in green. However, this template underpredicts the IR
emission by several orders of magnitudes.

\begin{figure}[tb]
{\centering
\includegraphics[width=8.8cm]{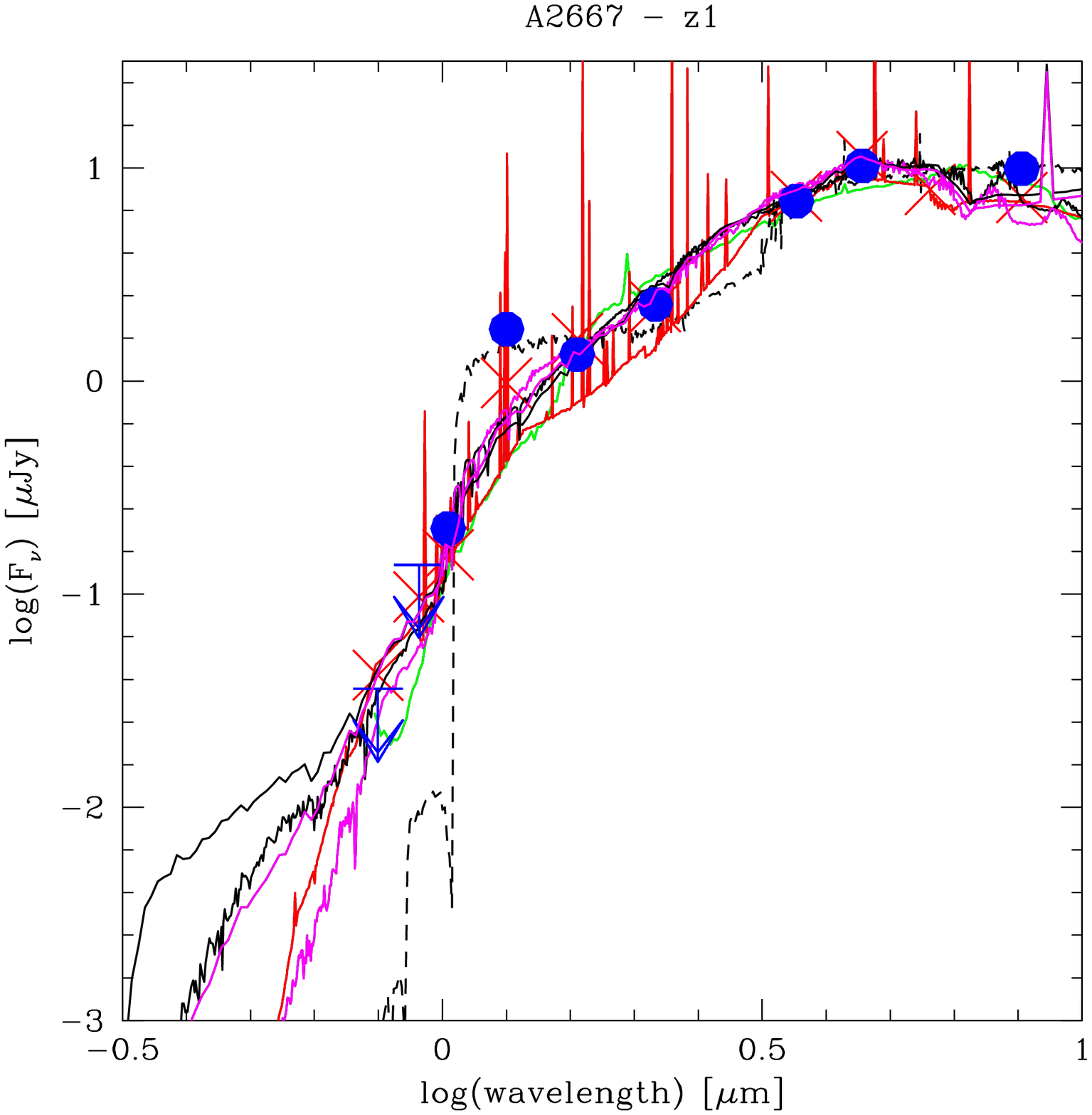}
}
\caption{Observed visible to near-IR part of the SED of z1 (blue circles) and best-fit models to this
spectral range:
using \citet{Polletta07} templates and no additional extinction (green line),
black lines: using \citet{2001ApJ...556..562C} and \citet{BC03} templates with variable
attenuation described by the Calzetti law,
magenta: same as black lines, but adoption the SMC extinction law,
red line (continuum) and red crosses (synthetic flux in filters): \citet{BC03} templates with nebular 
emission and SMC law,
black dashed: high redshift ($z=7.5$) template from \citet[][dashed line]{Laporte11}.
In all cases the best-fit redshift is $z \approx$ 1.5--1.7, except for Polletta's 
templates.
}
\label{fig_z1_detail}
\end{figure}

\subsection{Possible explanations for the strong SED break/very red spectrum}
As already seen, the common, observed spectral templates fail to reproduce
the steep, observed SED between the visible (I, z bands) and the near-IR (Y and J here), and they
predict a flux excess in the optical domain. 
What causes the sharp observed decrease of the flux between the near-IR and the optical
for these sources? 
%revds
{ The main difficulty provides from the fact that the largest spectral break known in galaxy
spectra is the Lyman break, whereas the typically observed Balmer (or ``4000 \AA'') break is 
smaller than that of our two galaxies.} 
We have examined various possibilities, but with no convincing answer.

For example, as shown in Fig.\ \ref{fig_z1_detail} for z1, adding variable extinction to the empirical 
templates allows one to diminish somewhat their optical flux excess. 
For the Chary \& Elbaz templates, the best-fit (to the domain shown here) is then with an additional extinction
$\Delta A_V=1.4$ for the Calzetti law. However, the templates are too smooth to reproduce the apparent break.
The flux at optical wavelengths can further be reduced by assuming a steeper
attenuation law than the Calzetti law adopted by default. With the SMC extinction law
by \citet{Prevot84} the best-fit is then for $\Delta A_V=1.2$. Overall, the same difficulty remains, however.
Figure \ref{fig_z1_detail} also shows the best-fit SED for \citet{BC03} templates with 
the SMC law. Although the steepest SED between the optical and the H-band, it falls short
in flux in the J band. 
For completeness we have also examined templates from the synthesis models of
\citet{Maraston06}. As expected, these models do not yield significantly different
fits in the blue part of the rest-frame optical spectrum.

As already mentioned by \citet{Laporte11}  we have also attempted to fit the SEDs 
with our models including nebular lines
\citep[see][]{SdB09}. Indeed, in this case the best-fit is found at $z \sim 1.5$ such that
the [\oiii ] $\lambda\lambda$ 4959,5007 lines, and H$\beta$ boost somewhat the J-band flux,
and H$\alpha$ the H-band to a lesser extent, contributing thus to the flux decrement between
J and Y. However, this solution requires also a very large attenuation ($A_V \sim 4.0$ for 
the SMC law) to reproduce the steeply rising SED towards longer wavelengths
\footnote{ Our model assumes identical attenuation for the continuum and nebular
lines, as in \cite{SdB09}}.
Although to the best of our knowledge objects with such red SEDs and strong emission lines 
are not known, this extreme explanation should be easy to test with
spectroscopic observations.

Finally, could composite populations not taken into account by our models 
help to explain the observed SED? Certainly the theoretical SED models may suffer
from this simplification. However, we do not see how this could help to resolve 
the problem with the large observed  spectral break, since a superposition
of individual simple stellar populations (not capable of reproducing this observation)
can only average out spectral features.
We conclude that we have no convincing explanation for the observed rapid drop of the
observed SED of our two sources.

%%%%%%%%%%%%%%%%%%%%%%%%%%%%%%%%%%%%%%%%%%%%%%%%%%%%%%%%%%%%%%%%%%%%%%%%%%%%%%%
\section{Discussion}\label{sec:comparison}

\subsection{Comparison with other objects in the literature}
How do our sources compare with other known galaxies and what is their nature?

By design our sources are near-IR selected, optical drop-out sources, i.e.\ sources with a very 
red color between the $J$ and $z$ band and/or between $Y$ and $J$. Our sources can therefore 
be compared to those selected by \citet{Capak11} from the COSMOS survey. From their Fig.\ 13 we 
note that with (I-J) $>$ 5.4 and 4.4 and (z-J) $>$ 3.9 and 3.1 for z1 and Y5, both sources show extreme 
(very red) optical to near-IR colors, when compared to other low redshift galaxies with red (J-z) colors.
 z1 and Y5 are also similar to the z-dropout galaxy HUDF-J2 identified by
\citet{2005ApJ...635..832M} as a $z \approx 6.5$ post-starburst galaxy candidate, but later shown 
to be most likely at $z \sim$ 1.8--2.5 interloper by 
\citet{Schaerer07ero,2007MNRAS.376.1054D,2007ApJ...665..257C}.
Although similar in several respects, HUDF-J2 shows a more monotonously rising SED
between z, J, and H than our objects exhibiting a ``sharper'' break.
The colors and fluxes of our sources are also very similar to those of the lensed optical drop-out galaxy 
\#2 found behind the cluster Abell 1835, identified with the $z=2.93$ sub-mm galaxy
SMMJ14009+0252 \citep[see][ and references therein]{Schaerer07ero,2009ApJ...705L..45W}.

Our sources are obviously also characterised by a large IR to optical flux ratio, a criterium which
has been used by various authors. 
For example, \citet{2007A&A...470...21R} have studied IRAC 3.6 \micron\
selected sources undetected in deep optical {\em HST} images. The (K-3.6) and (z-3.6) colors of z1 and Y5
are comparable to their sources; the main difference seems to be in (H-K), where our sources
are bluer than those of \citet{2007A&A...470...21R}.  z1 and Y5 appear to be related to the
very dusty $z \sim$ 2--3 sources from this study.
\citet{2008ApJ...672...94F} and \citet{2008ApJ...677..943D} have examined 24 \micron\ selected sources with very red
colors between 24 \micron\ and the R band ($S(24)/S(R) \ga$ 1000). They concluded that
the bulk of these sources are very luminous strongly dust-obscured galaxies (referred to as DOGs) 
at $z \sim 2$, powered by AGN and/or by starbursts. Using the I band as a proxy for R, we obtain
a flux ratio $S(24)/S(R) \ga 20000$ for z1, an extremely high flux ratio compared to the
other samples. From the SED of Y5 (cf.\ Fig.\ \ref{fig_y5}), we also expect this galaxy to 
show a high 24 \micron\ to optical ratio. 
According to the source density from  \citet{2008ApJ...677..943D} we would have expected $\sim$ 4 
 strongly dust-obscured galaxies down to $\sim$ 0.3 mJy at 24 \micron\ in our 45 arcmin$^2$ field.
The depth of our MIPS observations being similar, this value is comparable to our source density,
although our selection is different.
\citet{2008ApJ...689..127P} have also compared DOGs and sub-mm galaxies (SMGs) in the same R-K-24 \micron\
colors, showing that $\sim$ 30\% of SMGs satisfy the DOG criteria, the remainder showing less
extreme (i.e.\ bluer) colors.
This confirms that the SEDs of our sources are comparable to a subset of SMGs with the most
extreme optical to IR/sub-mm colors, as already seen above (Sect.\ \ref{s_models}).
 Our optical data is not deep enough to ascertain whether z1 and Y5 fulfill the usual criteria for 
extremely red objects (EROs), $(R-K) \ga 5.6$ in Vega magnitudes, at least as estimated from (I-Ks).

Among known sub-mm galaxies, one source, GN10 or GOODS 850-5, stands out as having particularly extreme
IR/sub-mm to visible/near-IR properties, similar to our two galaxies. 
Indeed, this source is undetected down to $\sim 0.01 \mu$Jy (1 $\sigma$) in the visible, shows fluxes of 
$\sim 1-5 \mu$Jy in the IRAC bands (3.6--8 \micron), and peaks at $\sim 10-20$ mJy around 1 mm
\citep{2004ApJ...613..655W,2009ApJ...690..319W,2009ApJ...695L.176D}, quite comparable to z1 and Y5.
However, GN10 remains undetected even at JHK \citep{2009ApJ...690..319W}, which can be explained by its higher
redshift ($z \sim 4$), recently confirmed from CO spectroscopy \citep{2009ApJ...695L.176D}.
The observed spectral break of GN10 found between 3.6 and 2.2 \micron\ and other
considerations \citep{2009ApJ...690..319W,2009ApJ...695L.176D} suggest a very high attenuation of $A_V \ga$ 4.5--5
for this source, or at least for the star-forming part of it, if hosting multiple components.
If we assume constant star-formation as \citet{2009ApJ...695L.176D} for their SED modeling, we would infer
$A_V \sim$ 3 (7.8) mag for z1 (Y5). Comparing the infrared-derived SFR with the upper limits
in the rest-frame UV domain, we can also estimate the attenuation of our sources. Adopting
the I-band flux as a constraint for the UV flux at $\sim$ 2300-2500 \AA\ and using the
\citet{Kennicutt98} calibration, we obtain $A_V \ga 4$ mag for both sources. 

 Sub-mm galaxies are also known to exhibit very strong attenuation. 
For example, sources with Balmer decrement measurements indicate $A_V \sim$ 1--3 
\citep{2004ApJ...617...64S,2006ApJ...651..713T}, and from SED fits \citep{2004ApJ...617...64S} estimate
$A_V = 3.0 \pm 1.0$ for their sample. \citet{Wardlow10} find $A_V = 2.6 \pm 0.2$ from the 
median SED of sub-mm galaxies, but more extreme attenuations are found within the sub-mm 
galaxy samples \citep[see e.g.][]{2004ApJ...616...71S}. Also, the extremely red object, sub-mm detected
galaxy HR10 studied by \citet{1999ApJ...519..610D} shows $A_V \sim 4.5$, as inferred from
comparison of the IR and \ha\ star-formation rate.
In short, although higher than the typical/median value of sub-mm galaxies, the attenuation
of the sources z1 and Y5 is similar to that of some sub-mm galaxies, such as GN10 at $z=4.04$
and others at lower $z$. Our sources are also somewhat fainter, both in the rest frame near-IR
and in the IR than the typical sub-mm galaxies \citep[cf.][]{Wardlow10}.
Finally, our sources stand out by their large spectral break, which -- to the best of our knowledge
-- is unusual among intermediate redshift sources.

\subsection{The other high-z candidates of the survey}

{ Based on our FIR detections we have identified two potential interlopers among the ten high-z candidates discovered by \citet{Laporte11}. Most of the other candidates are in crowded regions where several sources emit in the FIR and are blended with each other, making any FIR measurement impossible. Two other candidates only seem to be clean from any contamination in the MIPS, PACS and SPIRE maps, namely Y3 and Y4. They remain undetected in all the bands. However, the FIR upper limits obtained do not allow us to discriminate between low and high redshift. On the other hand we can rule out that these sources are as extreme as z1 and Y5 in their IR/sub-mm to near-IR flux ratio, since they should otherwise clearly be detected in our Herschel images.}

%%%%%%%%%%%%%%%%%%%%%%%%%%%%%%%%%%%%%%%%%%%%%%%%%%%%%%%%%%%%%%%%%%%%%%%%%%%%%%%
\section{Conclusions}\label{sec:conclusion}

{ Analyzing the FIR SED of two high redshift dropout candidates we find that both galaxies are likely at $z\sim2$  rather than $z>7$. From the FIR point of view alone, both galaxies could be similar to ULIRGs or SMGs which are common at $z\sim2$. At $z>7$ the SEDs would imply extreme dust temperatures and luminosities. Fitting the global SEDs considering all the data available from visible to submm we estimate $z \sim$ 1.6--2.5. 

However, the optical/NIR part of both objects remains difficult to understand if at  $z\sim2$. They show a very strong and well defined spectral break (presumably the Balmer break), unusual among intermediate redshift sources.
The source z1 is extreme for both IR/visible and near-IR/optical colors. 
The source Y5 has a somewhat smaller drop between the near-IR/optical domain.
%revds The source Y5 is somewhat less extreme, but has still a very unusual near-IR/optical color.

We have examined several possible explanations for the extreme colors of these galaxies but none of them is entirely satisfactory.  More observations are required to understand their nature. Once a spectroscopic confirmation of their redshift is obtained we will be able to create new SED templates.

 Other extreme sources \citep[e.g. GN10][]{2004ApJ...613..655W,2005ApJ...632L..13L,2009ApJ...690..319W,2009ApJ...695L.176D} were found from  MIPS observations and submm searches. There could therefore be two complementary paths leading to similar, extreme sources.

Although spectroscopic confirmation is still required and all possible interlopers may not be detected in the FIR, this work shows that FIR observations can be very helpful to constrain the contamination  of high-z dropout searches by lower redshift galaxies.}

%%%%%%%%%%%%%%%%%%%%%%%%%%%%%%%%%%%%%%%%%%%%%%%%%%%%%%%%%%%%%%%%%%%%%%%%%%%%%%%
\begin{acknowledgements}
We are grateful to the referee for constructive comments and suggestions that helped to improve the paper.
We thank the APEX staff for their aid in carrying out the
observations. APEX is operated by the Max-Planck-Institut
f\"ur Radioastronomie, the European Southern Observatory, and
the Onsala Space Observatory
This work received support from the french Agence Nationale de la Recherche under the reference ANR-09-BLAN-0234.
The work of DS and MZ is supported by the Swiss National Science Foundation. Support from ISSI (International Space Science Institute) in Bern for an
“International Team” is gratefully acknowledged. 
IRS, RJI, and AWB acknowledge support from STFC. 
\end{acknowledgements}

%%%%%%%%%%%%%%%%%%%%%%%%%%%%%%%%%%%%%%%%%%%%%%%%%%%%%%%%%%%%%%%%%%%%%%%%%%%%%%%
\bibliographystyle{aa}
\bibliography{hls_highz}
%\bibliography{hls_highz}

\end{document}